\documentclass[conference, 10pt, letterpaper]{IEEEtran}
\usepackage{amsfonts,cite,graphicx}

\newtheorem{theorem}{Theorem}

\begin{document}
\title{Sum Rate Maximization using Linear Precoding and Decoding in the
Multiuser MIMO Downlink}
\author{
\IEEEauthorblockN{Adam~J.~Tenenbaum and Raviraj~S.~Adve}
\IEEEauthorblockA{
Dept. of Electrical and Computer Engineering, University of Toronto\\
10 King's College Road, Toronto, Ontario, M5S 3G4, Canada\\
Email: \texttt{\{adam,rsadve\}@comm.utoronto.ca}}}

\maketitle
\begin{abstract}
We propose an algorithm to maximize the instantaneous sum data rate
transmitted by a base station in the downlink of a multiuser
multiple-input, multiple-output system. The transmitter and the receivers
may each be equipped with multiple antennas and each user may receive
more than one data stream. We show that maximizing the sum rate is
closely linked to minimizing the product of mean squared errors (PMSE).
The algorithm employs an uplink/downlink duality to iteratively design
transmit-receive linear precoders, decoders, and power allocations that
minimize the PMSE for all data streams under a sum power constraint.
Numerical simulations illustrate the effectiveness of the algorithm and
support the use of the PMSE criterion in maximizing the overall
instantaneous data rate.
\end{abstract}

\section{Introduction}
Multiple-input multiple-output (MIMO) systems continue to be an important
theme in wireless communications research. MIMO technology improves
reliability and/or increases the data rate of wireless transmission.
These performance improvements are achieved by exploiting the spatial
dimension using an antenna array at the transmitter and/or at the receiver.
A relatively recent theme has been MIMO systems enabling {\em multiuser}
communications in the downlink -- a single base station communicating with
multiple users.

Much of the existing work on multiuser MIMO systems focuses on minimizing
the sum of mean squared errors (SMSE) between the transmitted and
received signals under a sum power constraint~\cite{TA04,
SS05,SSJB05,KTA06,SB04}. A common theme to most of this work is the use
of an MSE uplink-downlink duality introduced in~\cite{SB04}. The work
in~\cite{MJHU06} provides a comprehensive review of the available work in
this area including an alternative algorithmic approach to this problem.
With its focus on SMSE, this body of work deals exclusively with
maximizing reliability at a fixed data rate. In particular, when one
considers the behaviour of the power allocation step in the SMSE
solutions, an ``inverse waterfilling'' type of solution may arise. When
starting at an optimum point for a fixed power allocation where data
streams have unequal powers, incremental power that is allocated to the
system will be assigned to the \emph{worst} of the active data streams.
This is required under the SMSE criterion, as the worst data stream's MSE
dominates the average (and thus, the sum) MSE.

This exclusive focus on minimizing error rate appears to hold contrary to
an important motivation in deploying MIMO systems: increasing data rate.
The problem of maximizing data rate has been studied in depth in
information theory, where sum capacity is attained by maximizing mutual
information.  In contrast to SMSE minimization, information theoretical
approaches apply a waterfilling strategy to assign available incremental
power to the \emph{best} data stream~\cite{SBJ03,VT03,YC04,JRVJG05}.
Unfortunately, the sum-capacity precoding strategy~\cite{Costa83} can not
be realized practically, and even suboptimal approximations (e.g.~those
employing Tomlinson-Harashima precoding~\cite{WFVH04}) require nonlinear
precoding, user ordering, and incur additional complexity.
Orthogonalization based methods using zero-forcing and block
diagonalization allow for a simple formulation of the sum
capacity~\cite{LJ06}, but the resulting constraint on the number of
receive antennas can severely restrict the possibility of receive
diversity and/or the associated increase in sum capacity.  Several papers
have looked at the general problem of maximizing sum capacity using
linear precoding for the multiuser downlink with single antenna
receivers~\cite{SJHU06,SVH06,BTC06}, but only recently has work been
performed on the case of multiple receive antennas~\cite{CTJL07}.

One important connection that we formulate in this paper is the
relationship between the sum capacity and the product of mean squared
errors (PMSE).  In the single-user multicarrier case, minimizing the PMSE
is equivalent to minimizing the determinant of the MSE matrix and thus is
also equivalent to maximizing the mutual information~\cite{PCL03}.  This
equivalence can also be seen in the relationship developed between
minimum MSE (MMSE) and mutual information in~\cite{GSV05}.  The existence
of these relationships motivates us to consider a PMSE minimizing
solution for the multiuser downlink to maximize the sum data rate over
multiple users, possibly with multiple data streams per user, given a
maximum allowable transmission power and constraints on the error rate of
each stream.

Information theoretical results for achieving sum capacity provide an
upper bound for achievable performance; however, a practical system
cannot use Gaussian codebooks in the design of its transmit
constellations. With this in mind, we evaluate the performance of our
PMSE minimizing linear precoder under adaptive PSK modulation.  The
resulting algorithm attempts to maximize the sum data rate, under PSK
modulation, with a constraint on the bit error rate of each data stream.
To our knowledge, this form of sum rate maximization (as opposed to that
performed in a purely information theoretic sense) has not been attempted
before.

The remainder of this paper is organized as follows.
Section~\ref{section:model} states the assumptions made and describes the
system model used.  Section~\ref{section:pmse_mot} investigates the
motivation for using the product of MSEs as an optimization criterion,
and Section~\ref{section:pmse_alg} proposes an optimization algorithm to
minimize the PMSE under a sum power constraint. Results of simulations
testing the efficacy of the proposed approach are presented in
Section~\ref{section:sims}. Finally, we draw conclusions in
Section~\ref{section:conc}.

\section{System Model\label{section:model}}
\begin{figure}
\begin{center}
\includegraphics[width=3.45in]{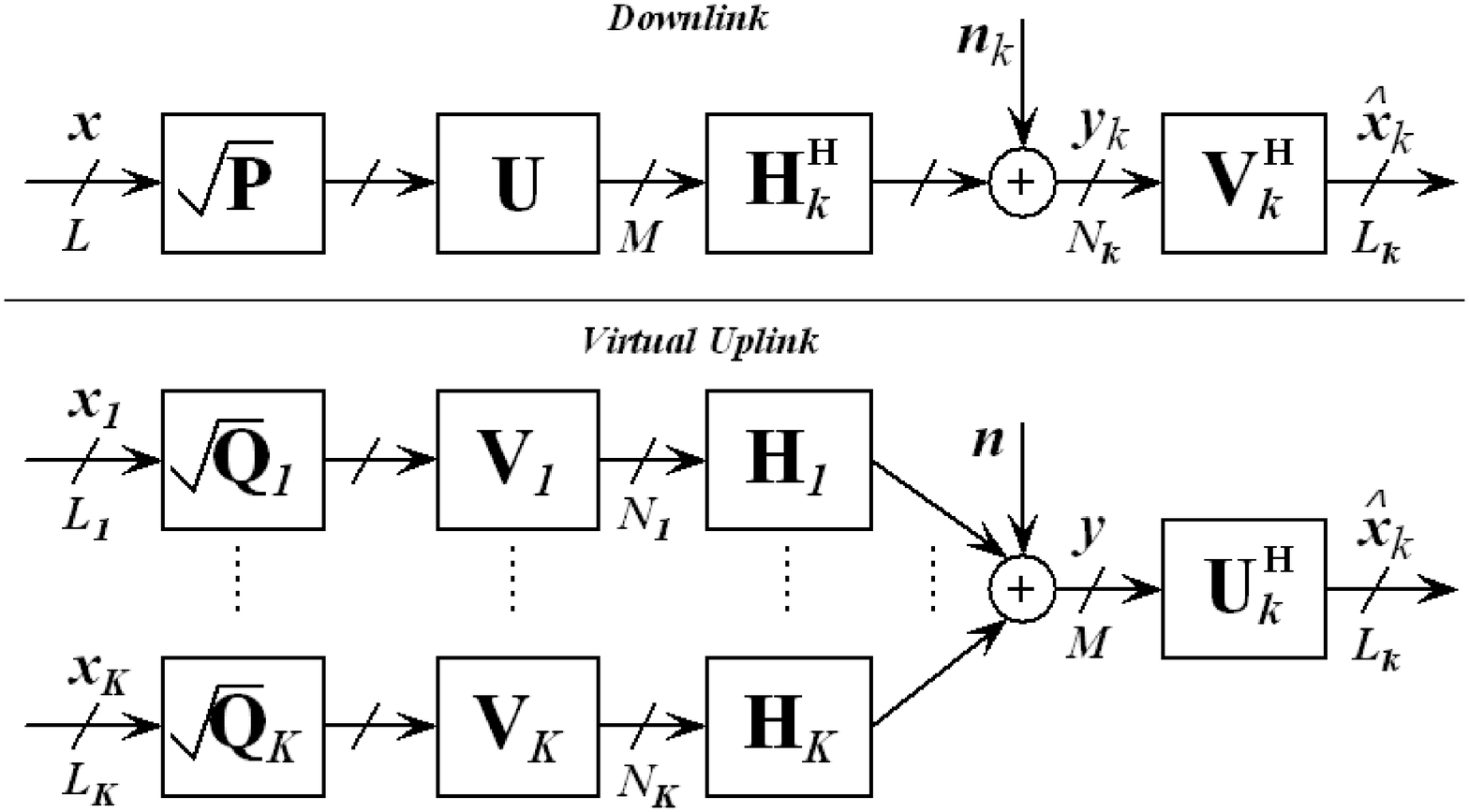}
\end{center}
\caption{Processing for user $k$ in downlink and virtual uplink.}\label{fig:scheme}
\end{figure}

The system under consideration, illustrated in Fig.~\ref{fig:scheme},
comprises a base station with $M$ antennas transmitting to $K$
decentralized users. User $k$ is equipped with $N_{k}$ antennas and
receives $L_k$ data streams from the base station. Thus, we have $M$
transmit antennas transmitting a total of $L=\sum_{k=1}^{K}L_{k}$ symbols
to $K$ users, who together have a total of $N=\sum_{k=1}^{K}N_{k}$
receive antennas. The data symbols for user $k$ are collected in the data
vector $\mathbf{x}_{k}=\left[x_{k1}, x_{k2}, \ldots, x_{kL_{k}}\right]^T$
and the overall data vector is $\mathbf{x} = \left[\mathbf{x}_1^T,
\mathbf{x}_2^T, \ldots, \mathbf{x}_K^T \right]^T$. We focus here on
linear processing at the transmitter and receiver. Hence, to ensure
resolvability we require $L \le M$ and $L_k \le N_k$, $\forall k$.

User $k$'s data streams are processed by the $M\times L_k$ transmit
filter $\mathbf{U}_{k} = \left[ \mathbf{u}_{k1}, \ldots,
\mathbf{u}_{kL_k}\right]$ before being transmitted over the $M$ antennas.
Each $\mathbf{u}_{kj}$ is the precoder for stream $j$ of user $k$, and
has unit power ($\|\mathbf{u}_{kj}\|_2 = 1$, where $\|\cdot\|_2$ is the
Euclidean norm operator).  These individual precoders together form the
$M\times L$ global transmitter precoder matrix $\mathbf{U} =
\left[\mathbf{U}_{1}, \mathbf{U}_{2}, \ldots, \mathbf{U}_{K}\right]$. Let
$p_{kj}$ be the power allocated to stream $j$ of user  $k$ and the
downlink transmit power vector for user $k$ be
$\mathbf{p}_{k}=\left[p_{k1}, p_{k2}, \ldots, p_{kL_{k}} \right]^T$, with
$\mathbf{p}=\left[ \mathbf{p}_1^T, \ldots, \mathbf{p}_K^T \right]^T$.
Define $\mathbf{P}_{k}=diag\{\mathbf{p}_{k}\}$ and
$\mathbf{P}=diag\{\mathbf{p}\}$. The channel between the transmitter and
user $k$ is assumed flat and is represented by the $N_{k}\times M$ matrix
$\mathbf{H}_{k}^{H}$, where $(\cdot)^H$ indicates the conjugate transpose
operator.  The resulting $N\times M$ channel matrix is $ \mathbf{H}^H$,
with $ \mathbf{H}=\left[\mathbf{H}_{1},\,
\mathbf{H}_{2},\,\dots,\mathbf{H}_{K} \right].$ The transmitter is
assumed to know the channel perfectly.

Based on this model, user $k$ receives a length $N_{k}$ vector
\begin{eqnarray}\label{eqn:yk_rcvd}
\mathbf{y}_{k}=\mathbf{H}_{k}^{H}\mathbf{U}\sqrt{\mathbf{P}}\mathbf{x} +
\mathbf{n}_{k},
\end{eqnarray}
where $\mathbf{n}_k$ consists of the additive white Gaussian noise (AWGN)
at the user's receive antennas with power $\sigma^2$; that is,
$\mathbb{E}\left[\mathbf{n}_{k}\mathbf{n}_{k}^{H}\right]=\sigma^2
\mathbf{I}_{N_{k}}$, where $\mathbb{E}\left[\cdot\right]$ represents the
expectation operator. To estimate its $L_{k}$ symbols $\mathbf{x}_{k}$, user $k$
processes $\mathbf{y}_{k}$ with its $L_{k}\times N_{k}$ decoder matrix
$\mathbf{V}_{k}^{H}$ resulting in
\begin{eqnarray}\label{eqn:xest_rcvd_dn}
\hat{\mathbf{x}}_{k}^{DL}=\mathbf{V}_{k}^{H}\mathbf{H}_{k}^{H}\mathbf{U}
\sqrt{\mathbf{P}}\mathbf{x}+\mathbf{V}_{k}^{H}\mathbf{n}_{k},
\end{eqnarray}
where the superscript $^{DL}$ indicates the downlink.

The global receive filter $\mathbf{V}^{H}$ is a block diagonal
decoder matrix of dimension $L\times N$, $\mathbf{V} =
diag\left[\mathbf{V}_{1}, \, \mathbf{V}_{2}, \cdots,
\mathbf{V}_{K}\right]$, where each $\mathbf{V}_{k} =
\left[ \mathbf{v}_{k1}, \ldots,
\mathbf{v}_{kL_k}\right]$.

We make use of the dual virtual uplink, also illustrated in
Fig.~\ref{fig:scheme}, with the same channels between users and base
station. Let the uplink transmit power vector for user $k$ be
$\mathbf{q}_{k}=[q_{k1}, q_{k2}, \ldots, q_{kL_{k}}]^T$, with
$\mathbf{q}=[\mathbf{q}_{1}^T,\ldots,\mathbf{q}_{K}^T]^T$, and define
$\mathbf{Q}_{k}=diag\{\mathbf{q}_{k}\}$ and
$\mathbf{Q}=diag\{\mathbf{q}\}$.  The transmit and receive filters for
user $k$ become $\mathbf{V}_{k}$ and $\mathbf{U}_{k}^{H}$ respectively.
As in the downlink, the precoder for the virtual uplink contains columns
with unit norm; that is, $\|\mathbf{v}_{kj}\|_2 = 1$. The received vector
at the base station and the estimated symbol vector for user $k$ are
\begin{eqnarray}
\mathbf{y}&=&\sum_{i=1}^{K}\mathbf{H}_{i}\mathbf{V}_{i}\sqrt{\mathbf{Q}_{i}}
\mathbf{x}_{i}+\mathbf{n},
\label{eqn:y_rcvd} \\
\hat{\mathbf{x}}_{k}^{UL}&=&\sum_{i=1}^{K}\mathbf{U}_{k}^{H}\mathbf{H}_{i}
\mathbf{V}_{i}\sqrt{\mathbf{Q}_{i}}\mathbf{x}_{i}+
                        \mathbf{U}_{k}^{H}\mathbf{n}.
\label{eqn:xest_rcvd_up}
\end{eqnarray}
The noise term, $\mathbf{n}$, is again AWGN with
$\mathbb{E}\left[\mathbf{n}\mathbf{n}^{H}\right]=\sigma^{2}\mathbf{I}_{M}$.

We assume that the modulated data symbols $\mathbf{x}$ are drawn from a
PSK constellation where each data symbol $x_i$ has power
$|x_i|^2=1$.  Furthermore, the data symbols are independent so that
$\mathbb{E}\left[\mathbf{x}\mathbf{x}^{H}\right]=\mathbf{I}_{L}$.  Also, noise
and data are independent such that
$\mathbb{E}\left[\mathbf{x}_i \mathbf{n}^H\right]=\mathbf{0}$.  Finally,
we define a useful virtual uplink receive covariance matrix as
\begin{eqnarray}
\nonumber \mathbf{J} & = &
\mathbb{E}\left[\mathbf{y}\mathbf{y}^{H}\right]
= \sum_{k=1}^K \mathbf{H}_k\mathbf{V}_k\mathbf{Q}_k\mathbf{V}_k^H\mathbf{H}_k^H
                                                   + \sigma^2\mathbf{I}_M\\
& = & \mathbf{HVQV}^H\mathbf{H}^H + \sigma^2 \mathbf{I}_M.
\end{eqnarray}

\section{Product of Mean Squared Errors\label{section:pmse_mot}}
Information theoretical approaches characterize the sum
capacity of the multiuser MIMO downlink or broadcast channel (BC) by
solving the sum capacity of the equivalent uplink multiple access channel
(MAC) and applying a duality result~\cite{VT03, VJG04}.  The resulting
expression for the maximum sum rate in the $K$ user MAC is
\begin{eqnarray}
\nonumber R_{\mathrm{sum}} & = & \max_{\mathbf{\Sigma}_k} \log_2 \det
\left( \mathbf{I} + \frac{1}{\sigma^2}
      \sum_{k=1}^K \mathbf{H}_k \mathbf{\Sigma}_k \mathbf{H}_k^H \right) \\
\nonumber & & \mathrm{s.t.}\quad \mathbf{\Sigma}_k \succeq \mathbf{0}
                                                \hspace{0.55in} k = 1, \ldots, K\\
& & \sum_{k=1}^K \mathrm{tr}\left[ \mathbf{\Sigma}_k\right] \le
P_\mathrm{max},
\end{eqnarray}
where $\mathbf{\Sigma}_k \succeq \mathbf{0}$ indicates
$\mathbf{\Sigma}_k$ is a positive semi-definite transmit covariance
matrix for mobile user $k$ in the uplink.  In this section, we
approximate this sum rate in terms of each individual user's data rate.

Consider the signal to interference plus noise ratio (SINR) for stream $j$
belonging to user $k$ under the multiuser virtual uplink model defined in
Section~\ref{section:model}. Using (\ref{eqn:xest_rcvd_up}) and finding the
average received signal power
($\mathbb{E}\left[ |\hat{x}_{kj}|^2 \right]$) and
interference-plus-noise power corresponding to all other data streams and
AWGN, this stream achieves an SINR of
\begin{eqnarray}\label{eqn:SINRUL}
\gamma_{kj}^{UL} & = &
\frac{\mathbf{u}_{kj}^H\mathbf{H}_k\mathbf{v}_{kj}
q_{kj}\mathbf{v}_{kj}^H\mathbf{H}_k^H\mathbf{u}_{kj}}
{\mathbf{u}_{kj}^H\mathbf{J}_{kj}\mathbf{u}_{kj}},
\end{eqnarray}
where $\mathbf{J}_{kj} \doteq \mathbf{J} -
\mathbf{H}_k\mathbf{v}_{kj}q_{kj}\mathbf{v}_{kj}^H\mathbf{H}_k^H$ is the
virtual uplink interference-plus-noise receive covariance matrix. We
approximate the maximum rate for this stream as
\begin{eqnarray}\label{eqn:rateapprox}
R_{kj} & \approx & \log_2 \left( 1 + \gamma_{kj}^{UL} \right).
\end{eqnarray}
Under the central limit theorem, the interference-plus-noise becomes
Gaussian as the number of interfering streams increases, making the
approximation progressively better.

The goal of this work is to maximize the sum data rate subject to
constraints on the total available power. Using the approximation in
(\ref{eqn:rateapprox}), we formally state the optimization problem as:
\begin{eqnarray}\label{eqn:maxrates}
\nonumber\left(\mathbf{V}, \mathbf{q}\right) & = &
\arg\max_{\mathbf{V}, \mathbf{q}}
\sum_{k=1}^K\sum_{j=1}^{L_k}\log_2 \left( 1 + \gamma_{kj}^{UL} \right)\\
\nonumber & & \mathrm{s.t.}\quad \|\mathbf{v}_{kj}\|_2 = 1, \quad k = 1, \ldots, K\\
\nonumber & & \hspace{0.37in} q_{kj} \ge 0, \hspace{0.32in} j = 1, \ldots, L_k\\
& & \|\mathbf{q}\|_1 = \sum_{k=1}^K\sum_{j=1}^{L_k} q_{kj} \le
P_{\mathrm{max}},
\end{eqnarray}
where $\|\mathbf{q}\|_1$ is the 1-norm or the sum of all entries in
$\mathbf{q}$.

We can see from (\ref{eqn:SINRUL}) that the optimum linear receiver
$\mathbf{u}_{kj}$ does not depend on any other columns of $\mathbf{U}$;
furthermore, it is the solution to the generalized eigenproblem
\begin{eqnarray}
\mathbf{u}_{kj}^{\mathrm{opt}} & = &
\hat{e}_{\max}\left(\mathbf{H}_k\mathbf{v}_{kj}q_{kj}
\mathbf{v}_{kj}^H\mathbf{H}_k^H, \mathbf{J}_{kj} \right),
\end{eqnarray}
where $\hat{e}_{\max}(\mathbf{A},\mathbf{B})$ is the unit norm
eigenvector $\mathbf{x}$ corresponding to the largest eigenvalue
$\lambda$ in the generalized eigenproblem
$\mathbf{Ax}~=~\lambda\mathbf{Bx}$.  Within a normalizing factor, this
solution is equivalent to the MMSE receiver:
\begin{eqnarray}\label{eqn:u_mmse}
\mathbf{u}_{kj}^{\mathrm{opt}} & = &
\mathbf{J}^{-1}\mathbf{H}_k\mathbf{v}_{kj}\sqrt{q_{kj}}.
\end{eqnarray}

When using linear decoding with this MMSE receiver, the MSE matrix for the
virtual uplink is
\begin{eqnarray}
\mathbf{E} & = & \mathbb{E}\left[\left(\hat{\mathbf{x}} -
\mathbf{x}\right)\left(\hat{\mathbf{x}} - \mathbf{x}\right)^H\right] \nonumber \\
& = & \mathbf{I}_{L} - \sqrt{\mathbf{Q}}\mathbf{V}^H\mathbf{H}^H
\mathbf{J}^{-1}\mathbf{HV}\sqrt{\mathbf{Q}},
\end{eqnarray}
which follows from (\ref{eqn:u_mmse}) and the system model assumptions
stated in Section~\ref{section:model}.  Thus, the mean squared error for
user $k$'s $j^{\mathrm{th}}$ stream is
\begin{eqnarray}
\epsilon_{kj} & = & 1 - q_{kj} \mathbf{v}_{kj}^H\mathbf{H}_k^H
                             \mathbf{J}^{-1}\mathbf{H}_k\mathbf{v}_{kj}.
\end{eqnarray}

Now consider another optimization problem, minimizing the product of mean
squared errors (PMSE) under a sum power constraint,
\begin{eqnarray}\label{eqn:minpmse}
\nonumber\left(\mathbf{V}, \mathbf{q}\right) & = &
\arg\min_{\mathbf{V},\mathbf{q}} \prod_{k=1}^K\prod_{j=1}^{L_k}\epsilon_{kj}\\
\nonumber & & \mathrm{s.t.}\quad \|\mathbf{v}_{kj}\|_2 = 1, \quad k = 1, \ldots, K\\
\nonumber & & \hspace{0.37in} q_{kj} \ge 0, \hspace{0.32in} j = 1, \ldots, L_k\\
& & \|\mathbf{q}\|_1 = \sum_{k=1}^K\sum_{j=1}^{L_k} q_{kj} \le
P_{\mathrm{max}}.
\end{eqnarray}

\begin{theorem}
Under linear MMSE decoding at the base station, the optimization problems
defined by (\ref{eqn:maxrates}) and (\ref{eqn:minpmse}) are equivalent
problems.
\end{theorem}

\begin{IEEEproof}
Define the argument of the log term from (\ref{eqn:rateapprox}) as
$\alpha_{kj} \doteq 1 + \gamma_{kj}^{UL}$. Using (\ref{eqn:SINRUL}), we
can rewrite $\alpha_{kj}$ as
\begin{eqnarray}\label{eqn:OPSINR}
\alpha_{kj} & = &
\frac{\mathbf{u}_{kj}^H\mathbf{J}\mathbf{u}_{kj}}
{\mathbf{u}_{kj}^H\mathbf{J}\mathbf{u}_{kj} -
\mathbf{u}_{kj}^H\mathbf{H}_k\mathbf{v}_{kj}q_{kj}
\mathbf{v}_{kj}^H\mathbf{H}_k^H\mathbf{u}_{kj}}.
\end{eqnarray}
It follows that by using the MMSE receiver from (\ref{eqn:u_mmse}),
\begin{eqnarray}\label{eqn:invSINR}
\frac{1}{\alpha_{kj}} & = &
 1 - \frac{\mathbf{u}_{kj}^H\mathbf{H}_k
   \mathbf{v}_{kj}q_{kj}\mathbf{v}_{kj}^H\mathbf{H}_k^H
    \mathbf{u}_{kj}}{\mathbf{u}_{kj}^H\mathbf{J}
    \mathbf{u}_{kj}} \nonumber \\
& = & 1 - \frac{ \left(q_{kj}\mathbf{v}_{kj}^H\mathbf{H}_k^H
 \mathbf{J}^{-1}\mathbf{H}_k\mathbf{v}_{kj}\right)^2}{q_{kj}\mathbf{v}_{kj}^H
     \mathbf{H}_k^H\mathbf{J}^{-1}\mathbf{H}_k\mathbf{v}_{kj}} \nonumber \\
& = & 1 - q_{kj}\mathbf{v}_{kj}^H\mathbf{H}_k^H\mathbf{J}^{-1}
           \mathbf{H}_k\mathbf{v}_{kj} = \epsilon_{kj}.
\end{eqnarray}
Thus, under linear MMSE decoding, the MSE and SINR for stream $j$
belonging to user $k$ are related as
\begin{eqnarray}
\epsilon_{kj} = \frac{1}{1 + \gamma_{kj}^{UL}}. \label{eqn:MSEandSINR}
\end{eqnarray}
This relationship is similar to one shown for MMSE detection in
CDMA systems~\cite{MH94}.  By applying (\ref{eqn:MSEandSINR}) to
(\ref{eqn:maxrates}), we see that
\begin{equation}
\sum_{k=1}^K\sum_{j=1}^{L_k} \log_2 \left( 1 + \gamma_{kj}^{UL} \right )
= -\log_2 \left( \prod_{k=1}^K\prod_{j=1}^{L_k} \epsilon_{kj} \right ).
\end{equation}
Since the constraints on $\mathbf{v}_{kj}$ and $q_{kj}$ are identical in
(\ref{eqn:maxrates}) and (\ref{eqn:minpmse}), the problem of maximizing
sum rate in (\ref{eqn:maxrates}) is therefore equivalent to minimizing
the PMSE in (\ref{eqn:minpmse}).
\end{IEEEproof}

\section{PMSE Minimization Algorithm\label{section:pmse_alg}}
With the motivation of Section~\ref{section:pmse_mot} in mind, we now
develop an algorithm to minimize the product of mean squared errors.  The
algorithm draws on previous work in minimizing the sum MSE~\cite{SSJB05,KTA06}.
It operates by iteratively obtaining the downlink precoder matrix
$\mathbf{U}$ and power allocations $\mathbf{p}$ and the virtual uplink
precoder matrix $\mathbf{V}$ and power allocations $\mathbf{q}$. Each
step minimizes the objective function by modifying one of these four
variables while leaving the remaining three fixed.

\subsection{Downlink Precoder}
For a fixed set of virtual uplink precoders $\mathbf{V}_k$ and power
allocation $\mathbf{q}$, the optimum virtual uplink decoder $\mathbf{U}$
is defined by (\ref{eqn:u_mmse}).  Each $\epsilon_{kj}$ is minimized
individually by this MMSE receiver, thereby also minimizing the product
of MSEs.  This $\mathbf{U}$ is normalized and used as the downlink precoder.

\subsection{Downlink Power Allocation}
The MSE duality derived in~\cite{SSJB05,KTA06} states that all achievable
MSEs in the \emph{uplink} for a given $\mathbf{U}$, $\mathbf{V}$, and
$\mathbf{q}$ (with sum power constraint $\|\mathbf{q}\|_1 \le P_{\max}$),
can also be achieved by a power allocation $\mathbf{p}$ in the
\emph{downlink} where $\|\mathbf{p}\|_1 \le P_{\max}$.

In order to calculate the power allocation $\mathbf{p}$, we apply the
following result from~\cite{KTA06}:
\begin{equation}
\mathbf{p}=\sigma^{2}(\mathbf{D}^{-1}-\mathbf{\Psi})^{-1}\mathbf{1},
\end{equation}
where $\mathbf{\Psi}$ is the $L\times L$ cross coupling matrix defined as
\begin{eqnarray}\label{eqn:coupling_mtrx}
[\mathbf{\Psi}]_{ij}=\left\{ \begin{array}{ll}
|\mathbf{\tilde{h}}_{i}^{H}\mathbf{u}_{j}|^{2}=|\mathbf{u}_{j}^{H}
\mathbf{\tilde{h}}_{i}|^{2} & \textrm{${i}\neq{j}$}\\
0 & \textrm{$i=j$}\end{array} \right.,
\end{eqnarray}
\begin{equation}
\nonumber \mathbf{D}=
\mathrm{diag}\left\{\frac{\gamma_{11}^{UL}}{|\mathbf{v}_{11}^{H}\mathbf{H}_{1}^{H}
\mathbf{u}_{11}|^2},\dots,
\frac{\gamma_{KL_{K}}^{UL}}{|\mathbf{v}_{KL_{K}}^{H}\mathbf{H}_{K}^{H}
\mathbf{u}_{KL_{K}}|^2}\right\},
\end{equation}
where $\mathbf{\tilde{H}}=\mathbf{H}\mathbf{V}=[\mathbf{\tilde{h}}_1,
\ldots,\mathbf{\tilde{h}}_{L}]$,
$\mathbf{U}=[\mathbf{u}_1,\ldots,\mathbf{u}_L]$, and $\mathbf{1}$ is the
all-ones vector of the required dimension.

\subsection{Virtual Uplink Precoder}
Given a fixed $\mathbf{U}$ and $\mathbf{p}$, the optimal decoders
$\mathbf{V}_k$ are the MMSE receivers:
\begin{eqnarray}
\mathbf{V}_k = \mathbf{J}_k^{-1} \mathbf{H}_k^H \mathbf{U}_k
\sqrt{\mathbf{P}_k}. \label{eqn:MSEdecoder}
\end{eqnarray}
In this equation, $\mathbf{J}_k \doteq \mathbf{H}_k^H \mathbf{UPU}^H\mathbf{H}_k +
\sigma^2 \mathbf{I}_{N_k}$ is the receive covariance matrix for user $k$.
The optimum virtual uplink precoders are then the normalized columns of
$\mathbf{V}_k$.

\subsection{Virtual Uplink Power Allocation}
The power allocation problem on the virtual uplink solves
(\ref{eqn:minpmse}) with a fixed matrix $\mathbf{V}$. In the minimization
of sum MSE, the corresponding step is a convex optimization
problem~\cite{KTA06}. Unfortunately, it is well accepted that the
power allocation subproblem in PMSE minimization (or equivalently, in
sum rate maximization) is non-convex~\cite{SJHU06,BTC06,CTJL07}.

We thus employ numerical techniques to solve the power allocation
subproblem, and use sequential quadratic programming (SQP)~\cite{BT95} to
minimize the PMSE. SQP solves successive approximations of a constrained
optimization problem and is guaranteed to converge to the optimum value
for convex problems; however, in the case of this non-convex optimization
problem, SQP can only guarantee convergence to a local minimum.  We note
that a similar approach was proposed in~\cite{CTJL07}, where iterations
of the the sum rate maximization problem are solved by local
approximations of the non-convex sum-rate function as a (convex)
geometric program~\cite{BV04}.

In summary, the PMSE minimization algorithm, motivated by a need to
maximize sum data rate, follows the same steps as the minimization of the
SMSE. The iterative algorithm keeps three of four parameters
($\mathbf{U},\mathbf{p},\mathbf{V},\mathbf{q}$) fixed at each step and
obtains the optimal value of the fourth.
Convergence of the overall algorithm to a local minimum is guaranteed since
the PMSE objective function is non-increasing at each of the four parameter
update steps.  Termination of the algorithm is determined by the
selection of the convergence threshold $\epsilon$.

While neither the overall problem (\ref{eqn:minpmse}) nor the power allocation
subproblem are believed to be convex, simulations suggest that changing the
initialization point has a minimal impact on the final solution; however,
initialization with the $\mathbf{U}$ and $\mathbf{p}$ found using the
SMSE algorithm in~\cite{KTA06} appears to reduce the number of iterations
required for convergence. A summary of our proposed algorithm can be found in
Table~\ref{table:pmse_algorithm}.
%
\begin{table}
\caption{Iterative PMSE minimization algorithm} 
\centering 
\begin{tabular}{l} 
\hline\hline\\
\textbf{Iteration:}\\
1- \textit{\ \ Downlink Precoder}\\
\qquad $\mathbf{\tilde{U}}_k =\mathbf{J}^{-1}\mathbf{H}_{k}^H\mathbf{V}_{k}
  \sqrt{\mathbf{Q}_{k}}$, \qquad $\mathbf{u}_{kj} = \frac{\mathbf{\tilde{u}}_{kj}}
     {\|\mathbf{\tilde{u}}_{kj}\|_2}$\\ \\
2- \textit{\ \ Downlink Power Allocation via MSE duality}\\
\qquad
$\mathbf{p}=\sigma^{2}(\mathbf{D}^{-1}-\mathbf{\Psi})^{-1}\mathbf{1}$\\
\\
3- \textit{\ \ Virtual Uplink Precoder}\\
\qquad $\mathbf{\tilde{V}}_k = \mathbf{J}^{-1}_{k}\mathbf{H}_{k}^{H}\mathbf{U}_{k}
  \sqrt{\mathbf{P}_{k}}$, \qquad $\mathbf{v}_{kj} = \frac{\mathbf{\tilde{v}}_{kj}}
            {\|\mathbf{\tilde{v}}_{kj}\|_2}$\\ \\
4- \textit{\ \ Virtual Uplink Power Allocation}\\
\qquad $\mathbf{q}=\arg\min_{\mathbf{q}}\prod_{k=1}^{K} \prod_{j=1}^{L_k}
 \epsilon_{kj} $, s.t. $q_{kj} \geq 0$, $\|\mathbf{q}\|_{1} \le P_{\max}$ \\
\\
5- \textit{\ \ Repeat 1--4 until
 $\left[\mathrm{PMSE}_{\mathrm{old}} - \mathrm{PMSE}_{\mathrm{new}}\right]/
 \mathrm{PMSE}_{\mathrm{old}} < \epsilon$}\\\\
\hline\hline
\end{tabular}
\label{table:pmse_algorithm} 
\end{table}

\section{Numerical Examples\label{section:sims}}
In this section, we present simulation results to illustrate the
performance of the proposed algorithms. In all cases, the fading channel
is modelled as flat and Rayleigh using a channel matrix $\mathbf{H}$
composed of independent and identically distributed samples of a complex
Gaussian process with zero mean and unit variance. The examples use a
maximum transmit power of $P_{\max}=1$; SNR is controlled by varying the
receiver noise power $\sigma^2$.  The transmitter is assumed to have
perfect knowledge of the channel matrix $\mathbf{H}$.

\subsection{Theoretical Performance}
\begin{figure}
\begin{center}
\includegraphics[width=3.45in]{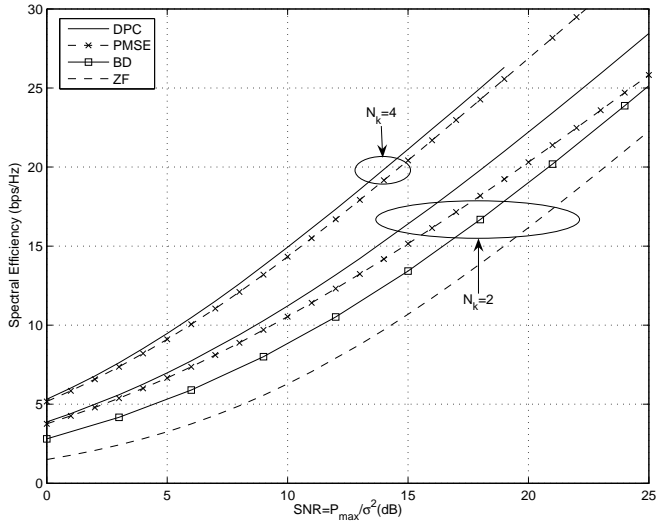}
\end{center}
\caption{PMSE vs. DPC and orthogonalization--based methods}\label{fig:PMSEDPC}
\end{figure}

First, we examine the information theoretical performance of the PMSE
algorithm proposed in Section~\ref{section:pmse_alg}.  That is, we
consider the spectral efficiency (measured in bps/Hz) that could be
achieved under ideal transmission by drawing transmit symbols from a
Gaussian codebook.  Figure~\ref{fig:PMSEDPC} illustrates how the proposed
scheme performs when compared to the sum capacity for the broadcast
channel (i.e. using dirty paper coding (DPC)~\cite{Costa83}) and to
traditional linear precoding methods based on channel orthogonalization
(i.e. block diagonalization (BD) and zero forcing (ZF)~\cite{LJ06}).
This simulation models a $K=2$ user system with $M=4$ transmit antennas
and $N_k=2$ or $N_k=4$ receive antennas per user.
The plot is generated using $30000$ channel realizations, with $5000$ data
symbols per channel realization, and the convergence threshold for the
PMSE algorithm is set as $\epsilon=10^{-6}$.

In Fig.~\ref{fig:PMSEDPC}, we see a slight divergence in the performance
of the PMSE algorithm from the theoretical DPC bound at higher SNR.  This
drop in spectral efficiency may be caused by the non-convexity of the
optimization problem, or it may suggest a fundamental gap between the
optimal DPC bound and the achievable sum capacity under linear precoding.
Nonetheless, the PMSE algorithm still maintains a higher spectral
efficiency than the orthogonalization based schemes for $N_k=2$.
Furthermore, the gap between the DPC bound and the PMSE precoder is only
0.6 dB for $N_k=4$, where BD and ZF schemes \emph{can not be applied} due
to constraints on the number of antennas.

\subsection{Performance Using Practical Modulation}
The precoder and decoder design algorithm in
Section~\ref{section:pmse_alg} is derived independently of modulation
depth, based on the assumption that transmitted symbols originate from
a unit-energy PSK constellation.  In this section, we consider two
approaches in selecting the modulation scheme to maximize data rate.

The \textit{naive} approach selects the largest PSK constellation of $b_{kj}$
bits per stream that satisfies a maximum bit error rate (BER) requirement
of $\beta_{kj}$.  The satisfaction of this constraint is determined using a
closed form BER approximation~\cite{CG01},
\begin{eqnarray}\label{eqn:mpsk}
\mathrm{BER}_{\mathrm{PSK}}(\gamma) \approx  c_1 \exp \left( \frac{-c_2
\gamma }{2^{c_3 b} - c_4}\right).
\end{eqnarray}
We apply the least aggressive of the bounds proposed in~\cite{CG01} by
using the values $c_1=0.25$,$c_2=8$,$c_3=1.94$, and $c_4=0$.  We note
that this approximation only holds for $b \ge 2$; as such, the following
exact expression should be used for BPSK:
\begin{eqnarray}\label{eqn:bpsk}
\mathrm{BER}_{\mathrm{BPSK}}(\gamma) \approx  \frac{1}{2}
\mathrm{erfc}\left(\sqrt{\gamma}\right).
\end{eqnarray}

The BPSK expression can be used as a test of feasibility for the
specified BER target; if the resulting BER under BPSK modulation is
higher than $\beta_{kj}$, then we have two options: either declare the
BER target infeasible, or transmit using the lowest modulation depth
available (i.e. BPSK).  In this work, we have elected to transmit using
BPSK whenever the PMSE stage has allocated power to the data stream.
Future work may consider either partial or complete non-transmission to
implement power saving while strictly achieving the desired BER target.

The naive approach is quite conservative in that there may be a large gap
between the BER requirement and BER achieved for each channel realization.  We
suggest a \textit{probabilistic} bit allocation scheme that switches between
$b_{kj}$ bits (as determined by the naive approach) and $b_{kj}+1$ bits with
probability
$p_{kj} = \left[\beta_{kj} - \mathrm{BER}_{b_{kj}}\right]/
\left[\mathrm{BER}_{b_{kj}+1} - \mathrm{BER}_{b_{kj}}\right]$.  This
modulation strategy may not be appropriate for systems requiring
instantaneous satisfaction of BER constraints; however, the
probabilistic method will still achieve the desired BER in the
long-term average over channel realizations.

\begin{figure}
\begin{center}
\includegraphics[width=3.45in]{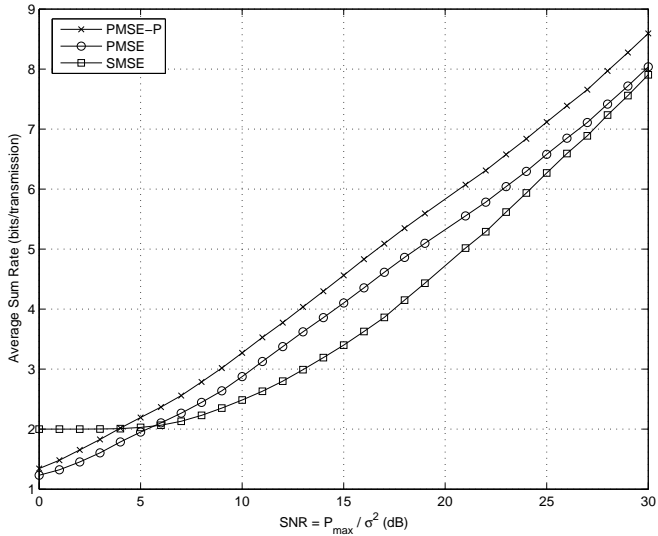}
\end{center}
\caption{Sum rate vs. SNR for user 1}\label{fig:sumrate}
\end{figure}

Figure~\ref{fig:sumrate} shows the sum rate achieved in the same system
configuration as described above ($K~=~2$, $M~=~4$, $N_k~=~2$) with the
additional required specification of $L_k=2$ data streams per user and
a target bit error rate of $\beta_{kj} = 10^{-2}$.  The plot
illustrates the average number of bits per transmission for user 1;
due to symmetry, the corresponding plot for user 2 is identical.  Note that in contrast to
Fig.~\ref{fig:PMSEDPC} (which shows the sum capacity under ideal
Gaussian coding), the sum rate in Fig.~\ref{fig:sumrate} is the
average number of bits transmitted in each realization using symbols
from a PSK constellation.

In Fig.~\ref{fig:sumrate}, we also consider using the naive PSK
modulation scheme for the PMSE precoder and the SMSE precoder designed
in~\cite{KTA06}.  Examination of this plot reveals that using the PMSE
criterion is justified at practical SNR values with improvements of
approximately one bit per transmission near 15 dB. Furthermore, using the
probabilistic modulation scheme (designated ``PMSE-P'') yields an
additional improvement of more than half a bit per transmission across
all SNR values.

\begin{figure}
\begin{center}
\includegraphics[width=3.45in]{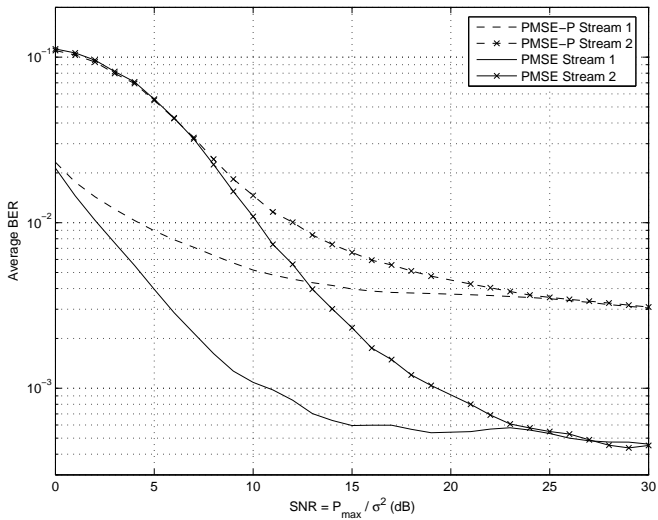}
\end{center}
\caption{BER vs. SNR for user 1}\label{fig:ber}
\end{figure}

In Fig.~\ref{fig:ber}, we plot average BER versus SNR for the same system
configuration as in Fig.~\ref{fig:sumrate}.  This plot illustrates how
the naive bit allocation algorithm attempts to achieve the target BER of
$10^{-2}$ for all data streams under PMSE, but also overshoots the
target, converging to a BER of approximately $5 \times 10^{-4}$.  This
can be attributed to the looseness of the BER bound, as discussed above.
In contrast, the probabilistic rate allocation algorithm not only
increases the rate, as shown in Fig.~\ref{fig:sumrate}, but also
converges to a BER that is much closer to the desired target BER. The
remaining gap between the actual BER achieved and the target BER can be
attributed to looseness in the approximations of~(\ref{eqn:mpsk})
and~(\ref{eqn:bpsk}).


\section{Conclusions}\label{section:conc}
In this paper, we have considered the problem of designing an iterative
method for maximizing bit rates in the multiuser MIMO downlink. Previous
work in the multiuser downlink has focused largely on added reliability
(minimizing SMSE), and not on maximizing the data rate. We have designed
a solution for a general MIMO system, where the number of users, base
station antennas, mobile antennas, and streams transmitted are only
constrained by resolvability of the data symbols. Our proposed solution
uses the SINR duality results from previous work in minimizing SMSE. The
product of the MSEs for all streams is minimized under a sum power
constraint; this is achieved by employing a known uplink-downlink duality
of MSEs. We also presented an adaptive modulation scheme to realize these
gains in rate in a practical system. The resulting SINR on each data
stream is then used to select an appropriate PSK constellation.
Simulations verify that significantly increased data rates can be
achieved while meeting given BER constraints.

\bibliographystyle{IEEEtran}
\bibliography{IEEEabrv,AJTArxiv}
\end{document}